# Ferromagnetic feature from Mn near room temperature in the fine particles of GdMn$_2$Ge$_2$ and TbMn$_2$Ge$_2$

**K. Mukherjee, Kartik K Iyer and E.V. Sampathkumaran**
*Tata Institute of Fundamental Research, Homi Bhabha Road, Colaba, Mumbai-400005, India*

**Abstract**

The magnetization behaviors of GdMn$_2$Ge$_2$ and TbMn$_2$Ge$_2$ in the bulk and in the fine particles (<1µm) obtained by high-energy ball-milling are compared. Pronounced modifications in the spontaneous, remanent and high-field magnetization in the fine particle form, attributable to Mn, are observed. The results indicate that the antiferromagnetism of Mn sub-lattice known for the bulk form in the range ~100 – 300 K gets weakened in favor of ferromagnetism in the fine particles. On the basis of this observation, we infer that there are other factors like size (and possibly defects) also play a role to decide the exact nature of magnetic ordering of Mn in this ternary family of materials, in contrasting to the traditionally held view that the basal plane Mn-Mn distance is the crucial controlling parameter.

(Version dated 2$^{nd}$ January 2010)

## 1. Introduction

Among the rare-earth (*R*) intermetallic compounds derived from the layered ThCr$_2$Si$_2$-type structure, the Mn-based compounds, RMn$_2$X$_2$ (X= Si, Ge), have been of great interest for the past four decades [1]. This is mainly due to the fact that the nature of the intralayer and interlayer magnetic couplings have been found to be extremely sensitive to the Mn-Mn separation ($d_a$) in the basal plane and often 'first-order' magnetic phase transitions and several magnetic anomalies as a function of temperature (*T*) have been observed in many compounds and pseudo-ternary alloys in this family [see, for instance, References 1-19]. It was believed [1] for a long time that the critical distance for a change in magnetic structure is somewhere in the range 2.84 to 2.87 Å. If $d_a$>2.87Å, the intralayer in-plane and interlayer couplings are antiferromagnetic and ferromagnetic respectively. If $d_a$ falls in the range 2.84 and 2.87 Å, intralayer in-plane coupling and interlayer coupling are antiferromagnetic. If $d_a$ falls below 2.84 Å, intralayer in-plane spin-component is essentially absent with antiferromagnetism between layers. Multiple transitions observed in some alloys are believed to occur if $d_a$ lies at the border regions due to variations in $d_a$ as a result of thermal contraction with decreasing temperature. For instance, in the case of the series under discussion, RMn$_2$Ge$_2$, one observes Mn-sublattice ferromagnetism near 300 K at the beginning of this series, whereas the heavier R members are found to be antiferromagnetic with Neél temperatures ($T_N$) above 300 K; for SmMn$_2$Ge$_2$, in the middle of the rare-earth series, the transitions at ~350 K (paramagnetism to ferromagnetism), ~150 K (ferromagnetism to antiferromagnetism), and ~100 K (antiferromagnetism to ferromagnetism) due to Mn sublattice have been observed. Recently, Duman et al [13] proposed that the critical distance is in fact different for silicides and gemanides and that the magnetism is decided by the exchange mechanism mediated by Si and Ge. It is thus clear that the origin of magnetic anomalies of this family is still under debate despite voluminous literature over several years.

The present investigation is primarily aimed at addressing how a reduction in particle size influences magnetism in this family, focusing on RMn$_2$Ge$_2$ series. Recently, we reported [20] that, in the case of a well-known non-magnetic heavy fermion compound, CeRu$_2$Si$_2$, the nano particles obtained by high-energy ball-milling are in fact magnetic below about 8 K, corresponding to the lattice-expansion effect on magnetism caused by partial La (for Ce) or Ge (for Si) substitution. With this clue, we wanted to explore whether antiferromagnetism seen in the room temperature range in the heavy R members in RMn$_2$Ge$_2$ can be driven towards ferromagnetism observed for light R members. With this primary motivation, we synthesized fine particles which are less than 1 μm in size for Gd and Tb members and studied the magnetization behavior. In the bulk form, the antiferromagnetic coupling among ferromagnetic Mn layers in GdMn$_2$Ge$_2$ sets in below 365 K and there is a first-order transition at about 95 K to ferrimagnetism with decreasing temperature following simultaneous Gd sublattice ordering [10, 14, 15, 16] and ferromagnetic alignment of all Mn planes. The corresponding $T_N$ for TbMn$_2$Ge$_2$ has been reported to be in the range 400 – 450 K, with the first-order ferromagnetic transition setting in at about 95 K [5, 17, 18, 19]. The present investigations reveal that a ferromagnetic component near room temperature indeed develops in the fine particle form, thereby suggesting that the particle size plays a role on Mn sublattice magnetic ordering. Additionally, we find that there are profound changes in the magnetization (*M*) at low temperatures following magnetic ordering of Gd and Tb sublattices, the transition temperatures of which remain unaffected in the fine particles. At this juncture, it is worthwhile to note that the ferromagnetism in $LaMn_2Ge_2$ is not altered in the nanoparticles [21] and hence it is clear that ferromagnetism is favoured in the nanoparticles in this family.

## 2.  Experimental details

The polycrystalline ingots of $GdMn_2Ge_2$ and $TbMn_2Ge_2$ were synthesized by repeated arc melting of stoichiometric amounts of constituent elements in an arc furnace in an atmosphere of argon. The alloys were then homogenized in an evacuated sealed quartz tube at 800 C for 7 days. The loss due to evaporation of Mn after first melting was compensated by adding corresponding amount of Mn before further melting. The materials thus obtained (called *B*) were milled in a medium of toluene for 5½ hrs in a planetary ball bill (Fritsch pulverisette-7 premium line) operating at a speed of 500 rpm. Zirconia vials and balls of 5mm diameter were used and the ball-to-material ratio was kept at 5:1. These vials and balls contain 96.4% of $ZrO_2$, and 3.2% of MgO and thus possible contamination from magnetic impurities due to these milling parts can be ignored. The phase purity of the molten ingot was ascertained by x-ray diffraction (Cu $K_α$) (see figure 1) and the homogeneity was further confirmed by a scanning electron microscopic pictures and energy dispersive x-ray analysis. A transmission electron microscope (TEM) (Tecnai 200 kV) was employed to infer particle size for the milled specimen (*N*) (after ultrasonification) and the pictures shown in figure 2 reveal the sizes of the particles lie in the micron range (<1 μm). Selected area diffraction patterns (figure 2) were also obtained with the same TEM to confirm that the fine particles indeed correspond to the desired phase. The dc $M(T)$ (1.8 – 330 K) and $M(H)$ (up to $H$= 50 kOe) measurements were performed with the help of a commercial magnetometer (Quantum Design).

## 3.  Results and discussion

*We first illustrate our findings in detail on $GdMn_2Ge_2$.*

The results of magnetization measurements as a function of temperature in 100 Oe and 5 kOe and as a function of magnetic field at selected temperatures for the bulk and the fine particles are compared in figure 3 and 4. For the bulk form of this compound, the $M/H$ (e.g., measured in a field of 5 kOe, shown in the inset of figure 3a in an expanded form, for the zero-field-cooled (ZFC) condition) increases with increasing *T* above ~ 150 K as a signature of the maximum for antiferromagnetic ordering setting in below 365 K as well-known in the literature. However, in the case of the specimen *N*, it is distinctly seen that the sign of the slope of this curve becomes negative above ~200 K; there is a broad peak below this temperature, which is cut off by the upturn as one approaches magnetic ordering temperature of Gd sublattice at 95 K. The magnitudes of $M/H$ also get enhanced with respect to those for *B*. In the data collected in low-fields (100 Oe, see figure 3b), the broad peak in $M/H$ for *N* is distinctly shifted to a higher temperature (~240 K) for the zero-field-cooled (from 330 K) condition of the specimen, attributable to low-field sensitivity of ferromagnetic feature coupled to a distribution of the peak temperature due to varying particle size. Thus, it is obvious that there is a significant change in the magnetic behavior of Mn sublattice in the fine particles with respect to that in bulk form. In support of this, the $M(T)$ curves obtained in 100 Oe for ZFC and field-cooled (FC) conditions (from 330 K) show a bifurcation near 300 K in *N*, whereas such a bifurcation sets in at the onset of Gd magnetic ordering only (below 95 K) in the case of bulk specimen. Such a bifurcation is a signature of pinning arising from a ferromagnetic component in this family [16]. It is worth noting that a small lattice expansion induced by a small substitution of Gd by Pr has been found to induce ferromagnetism around 270 K by Elerman et al [16]. However, there is no corresponding lattice expansion in the milled specimen (see figure 1). Therefore, a reduction in particle size and possibly defects due to milling presumably result in subtle modifications in electronic structure, which appears to lead to changes in magnetism mimicking that of lattice expansion.

Further evidence for enhancement of ferromagnetic component can be inferred from the comparison of *M*(*H*) curves as well (figure 4a). For instance, for *N*, at 120, 220 and 300 K, the variation of *M* with *H* for initial applications of field (< 5 kOe) is relatively steeper compared to that in *B*. In particular, this difference is dramatically seen at 120 K, as *M* varies linearly with *H* in the entire field range at this temperature for *B*; at higher fields, *M* for all the three representative temperatures varies linearly with *H* due to paramagnetism dominated by Gd ions. The magnetic moment obtained by linear extrapolation from high fields (>20 kOe) is quite significant and nearly the same at 120 and 220 K (about 0.2 $\mu_B$/formula unit), as though ferromagnetic component develops in *N*. Such a large magnitude of this extrapolated moment should arise from core-shell, as it is well-known in the literature that the corresponding value for any surface effect is two orders of magnitude smaller [22]. This large value also rules out any magnetic contamination from balls and vials, as these parts are nonmagnetic as mentioned earlier. In the case of *N*, we have observed hysteresis behavior as well in *M*(*H*) curves below 5 kOe (see insets of figure 4a) in conformity with a ferromagnetic component. This hysteresis is absent for the bulk specimen.

With respect to the behavior at lower temperatures, the Gd sub-lattice ordering temperature does not seem to be influenced in the fine particles, but the sharpness of the transition vanishes presumably due to strains induced by ball-milling, as inferred from *M*(*T*) plots (figure 3). The hysteretic nature of the plots of *M*(*H*) for *N*, as shown in figure 4b at three representative temperatures, are more prominent persisting even at high fields. The value of *M* increases gradually with increasing *H* without any evidence for saturation, for both *B* and *N* specimens. This finding implies that the ferrimagnetic character is retained in the fine particles as well following Gd sub-lattice ordering. However, non-zero spontaneous magnetization values (~0.9 and 0.28 $\mu_B$/formula-unit at 10 and 80 K respectively) apart from non-zero remnant magnetization (~1.2 and 0.36 $\mu_B$/formula-unit at 10 and 80 K respectively) are observed for *N* with increasing magnitude with decreasing temperature, thereby implying enhancement of net ferromagnetic component. As noted in figure 4b, the spontaneous magnetization values are negligible for *B*. For 1.8 K, the *M*(*H*) curve for *N* begins with the negative values which is sometimes known for two sub-lattice compounds like in ferrimagnets, though we are not able to explain why this feature is absent for the bulk specimen. If one is guided by the value of *M*, say, at 50 kOe, under the assumption of ferrimagnetic coupling between Gd and Mn layers, one infers that the moment on Mn at such high fields gets reduced in the ball-milled sample at low temperatures. For instance, at 1.8 K, this value is estimated to be about 2.8 and 1.7 $\mu_B$ for *B* and *N* respectively from the measured high-field $M$ value. Incidentally, this trend in high-field magnetization and remnant magnetization values are interestingly in good agreement with that we reported for ball-milled LaMn$_2$Ge$_2$ [21].

*We now focus on the behavior of the TbMn$_2$Ge$_2$.*

The results are shown in figures 5 and 6. The following features emphasized for the case of Gd compound are observed in the Tb case as well for the *N*-specimen in figure 5: (i) The bifurcation of low-field (100 Oe) ZFC-FC $M/H$ curves near room temperature as well as a broad peak (near 250 K) are apparent for the specimen *N*; (ii) Enhancement of $M/H$ values when the rare-earth sublattice is disordered in the high temperature range; (iii) broadening of the transition at ~100 K. The behavior of $M(H)$ (figure 6) above 100 K is also similar to the corresponding Gd case. That is, as demonstrated for 120, 275 and 330 K, there is a sharper increase of $M$ at low fields followed by a linear variation at high-fields for *N* in comparison with *B*; with decreasing temperature below 330 K, hysteresis nature of the curve gets enhanced for *N*, as apparent at 120 K; these are in support of a ferromagnetic component for *N* above 100 K. At low temperatures (<100 K), we see some differences with respect to the behavior of Gd compound. It is transparent from figure 6b that, though significant values of saturation moment obtained by linear extrapolation of high-field $M$ to zero-field and remnant magnetization are characteristic of both *B* and *N* specimens, these values are relatively reduced for *N*. However, the values of spontaneous magnetization are significantly higher for *N* at 5 and 75 K. This value at 1.8 K is nearly the same for both the forms, as though, for a small change in temperature at low temperatures, there is a subtle change in magnetism for fine particles. A striking difference with respect to the behavior in Gd case is that the high-field magnetization values for *N* are lower compared to those for *B*; in addition, the tendency towards saturation in *B*, for example, at 1.8 K, occurs at rather low fields, whereas, for *N*, the variation of $M$ with $H$ is more gradual as though very large fields are required to saturate magnetization for any meaningful inferences from high-field values. The change in the nature of the curves seems to imply that the Tb-sublattice in *N* might also have undergone a change in its magnetic structure following milling. A finding of note is that, in the bulk form, there is a major change of slope around 20 K in $M(T)$ plot (figure 5) as known earlier [19], which vanishes after milling; this might endorse profound changes in the magnetism due to Tb-sublattice at low temperatures.

## 4. Conclusion

On the basis of the comparison of magnetization behavior of bulk and fine particles of GdMn$_2$Ge$_2$ and TbMn$_2$Ge$_2$, we infer that, in the range 100-300 K, a ferromagnetic component tends to dominate in the fine particles of bulk antiferromagnets in this family. The magnetic behavior of Mn in this family of ternary compounds is thus shown to be to some extent dictated by the size of the particles and possibly defects introduced by milling, apart from the role of Mn-Mn separation and of the recently proposed role of the ligands [13].

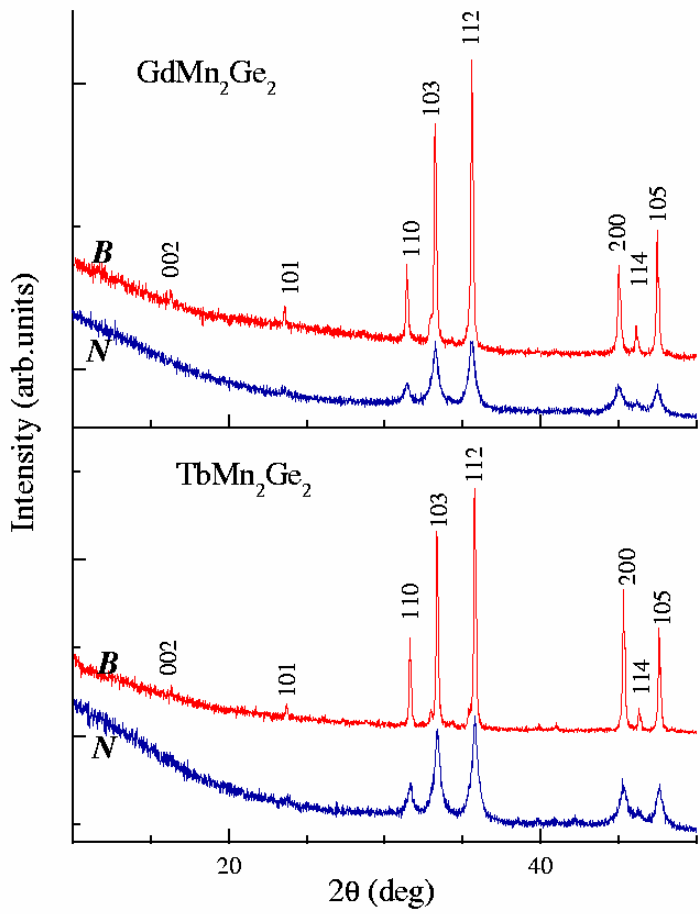

Figure 1:
X-ray diffraction patterns of bulk (**B**) and fine particles (**N**) for $GdMn_2Ge_2$ and $TbMn_2Ge_2$.

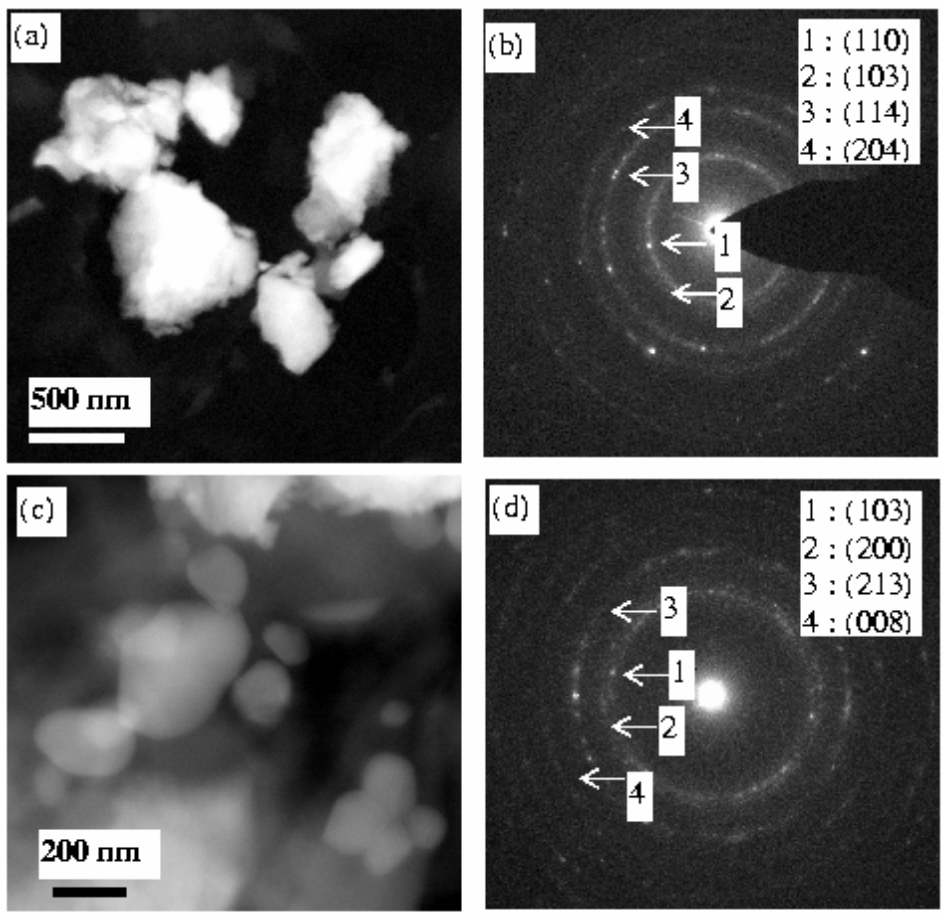

Figure 2:
TEM pictures and selected area diffraction patterns of $GdMn_2Ge_2$ (a and b) and $TbMn_2Ge_2$ (c and d).

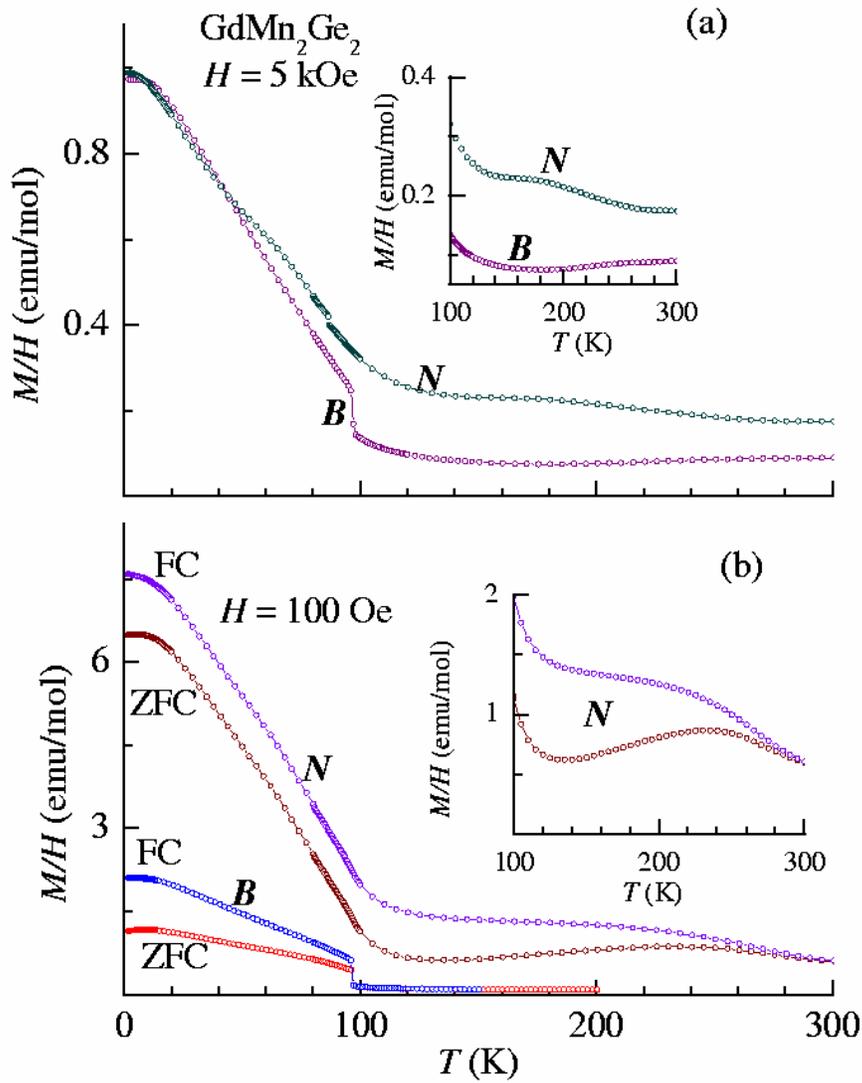

Figure 3:
Magnetization by magnetic field as a function of temperature for GdMn$_2$Ge$_2$ measured in a field of (***a***) 5 kOe for zero-field-cooled condition and (***b***) 100 Oe for zero-field-cooled and field-cooled conditions of the bulk and fine particles. In the insets, the data above 100 K are plotted in an expanded scale for ***N***. The lines through the data points serve as guides to the eyes.

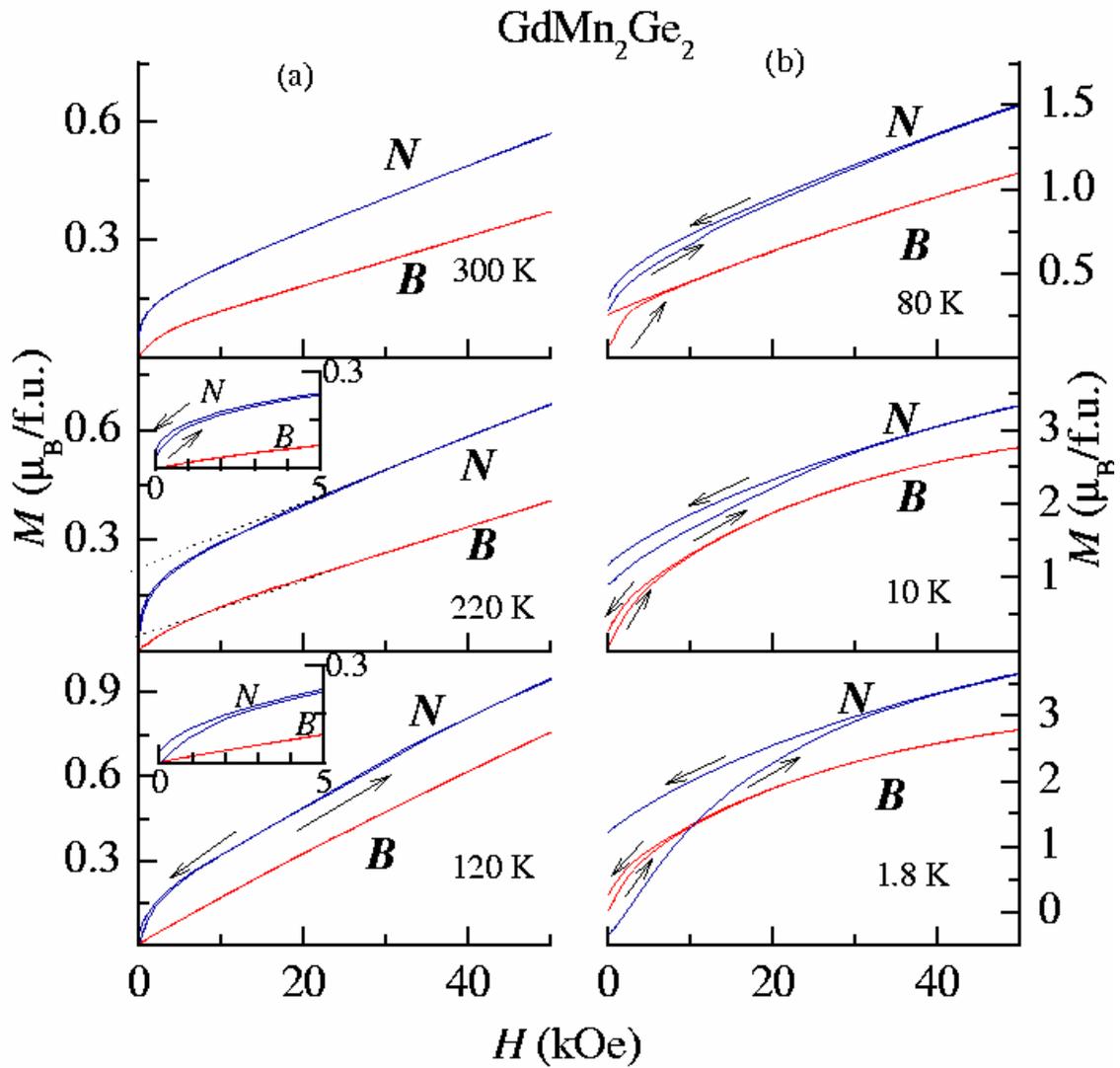

Figure 4:
Isothermal magnetization as a function of magnetic field at (***a***) 120, 220 and 300 K, and (***b***) 80, 10 and 1.8 K, for the bulk (***B***) and fine particles (***N***) of $GdMn_2Ge_2$. Dashed lines shown in some cases are extrapolations from high-field behavior. In the inset of (***a***), the data below 5 kOe is shown in an expanded form to highlight hysteresis for the fine particles; the curves for increasing and decreasing field cycles overlap for ***B***.

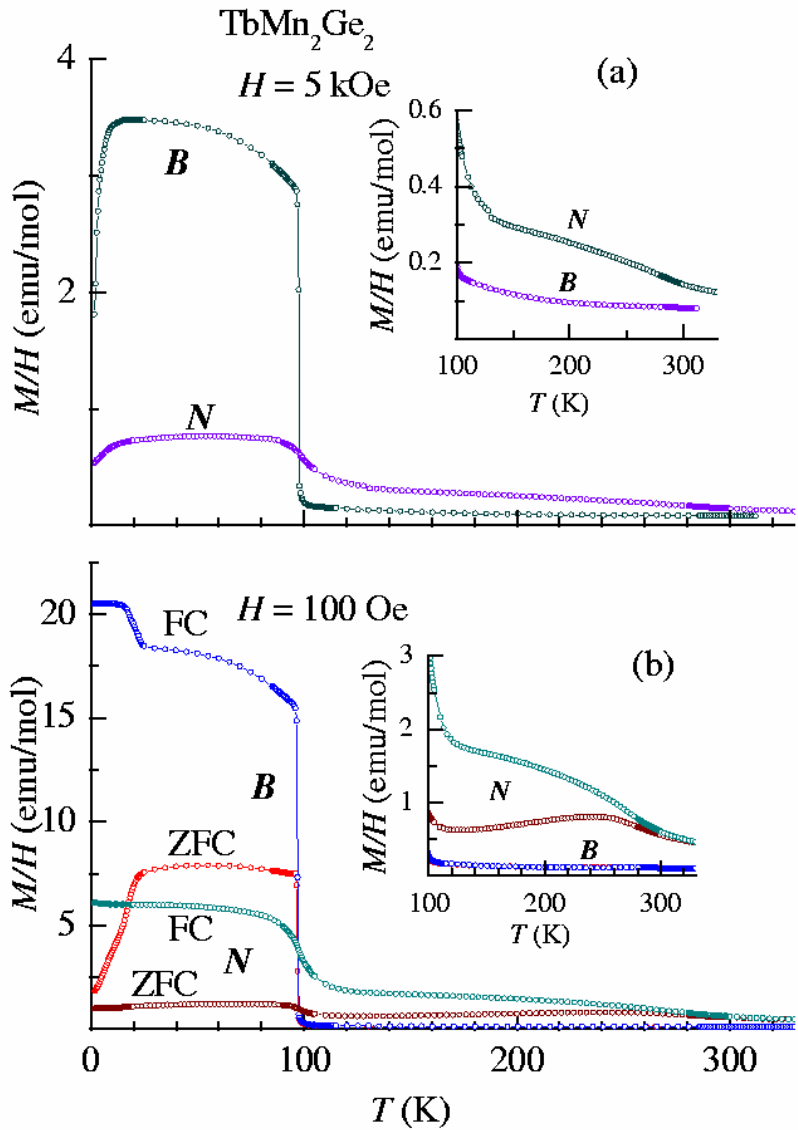

Figure 5:
Magnetization by magnetic field as a function of temperature for TbMn$_2$Ge$_2$ measured in a field of (***a***) 5 kOe for zero-field-cooled condition and (***b***) 100 Oe for zero-field-cooled and field-cooled conditions of the bulk and fine particles. In the insets, the data above 100 K are plotted in an expanded scale. The lines through the data points serve as guides to the eyes.

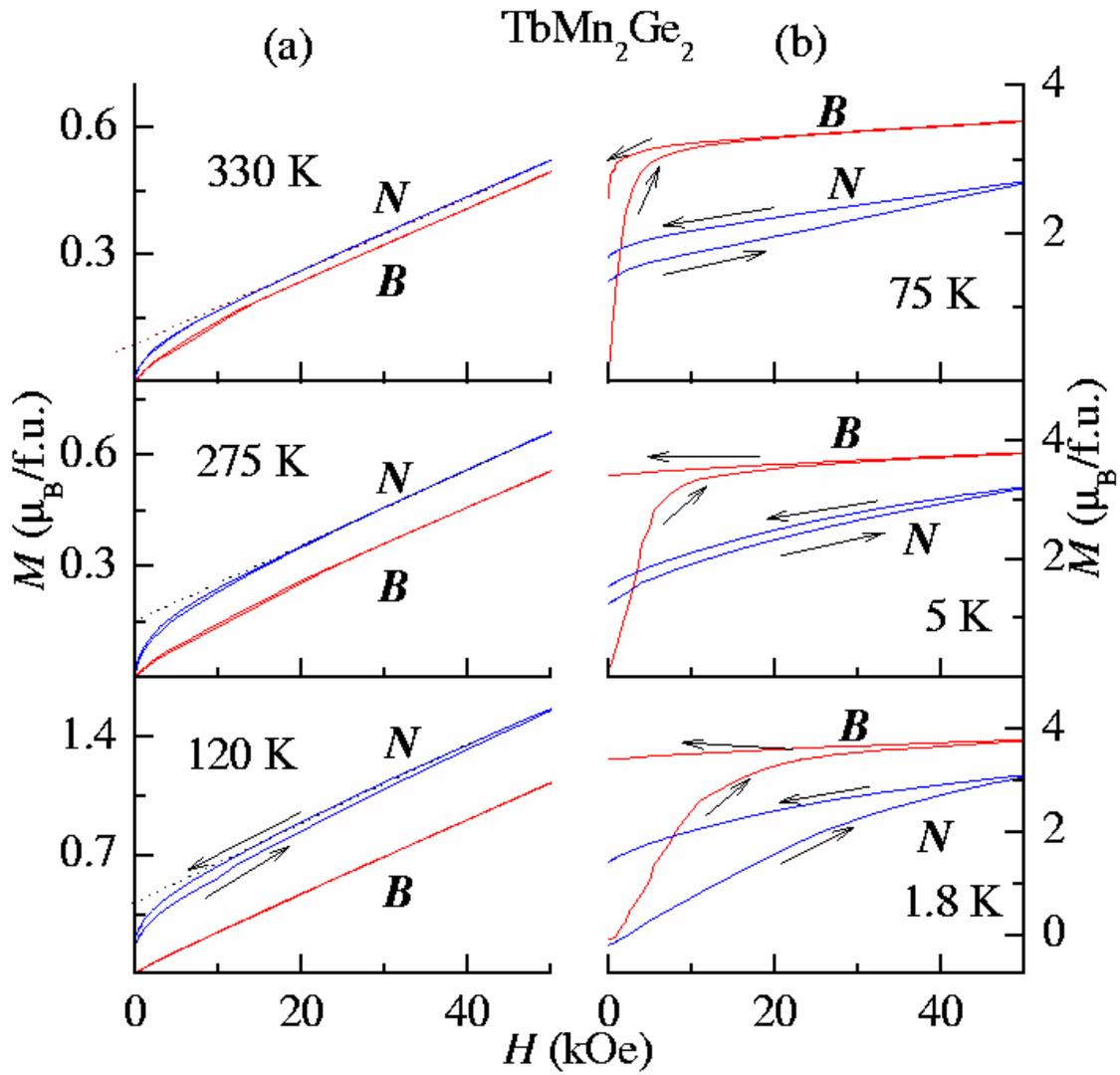

Figure 6:
Isothermal magnetization as a function of magnetic field at (*a*) 120, 275 and 330 K, and (*b*) 75, 5 and 1.8 K, for the bulk (*B*) and fine particles (*N*) for $TbMn_2Ge_2$. Dashed lines drawn in some cases are extrapolations from high-field behavior.